\documentclass{vgtc}                          

\graphicspath{{figures/}{pictures/}{images/}{./}} 

\usepackage{times}                     

\usepackage{tabu}                      
\usepackage{booktabs}                  
\usepackage{lipsum}                    
\usepackage{mwe}                       
\usepackage{enumitem}
\usepackage{mathptmx}                  
\usepackage{multirow}
\usepackage{amsmath}
\usepackage{booktabs}
\usepackage[table]{xcolor}

\onlineid{PacificVis 2026 VisNotes - submission 1155}

\vgtccategory{Research}

\vgtcinsertpkg

\title{Evaluating Visual Prompts with Eye-Tracking Data for MLLM-Based Human Activity Recognition}

\author{Jae Young Choi\thanks{e-mail: jaeyoungchoi@kaist.ac.kr}\\ %
        \scriptsize KAIST %
\and Seon Gyeom Kim\\ %
     \scriptsize KAIST %
\and Hyungjun Yoon\\ %
     \scriptsize KAIST
\and Taeckyung Lee\\
    \scriptsize KAIST
\and Donggun Lee\\
    \scriptsize KAIST
\and Jaeryung Chung\\
    \scriptsize KAIST
\and Jihyung Kil\\
    \scriptsize Adobe Research
\and Ryan Rossi\\
    \scriptsize Adobe Research
\and Sung-Ju Lee\thanks{Corresponding authors.}\\
    \scriptsize KAIST
\and Tak Yeon Lee\footnotemark[2]\\
    \scriptsize KAIST
     }
\teaser{
  \centering
  \vspace{-10pt}
  \includegraphics[width=\linewidth]{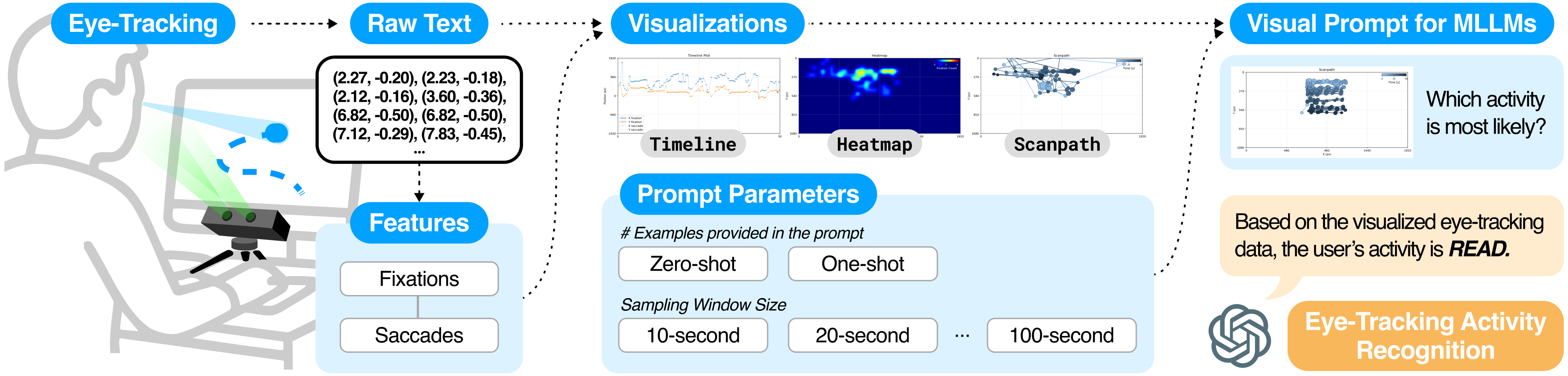}
  \vspace{-18pt}
  \caption{Overview of the study. We systematically explored visual prompting strategies with eye-tracking data by varying visualization types and prompt parameters (e.g., zero/one-shot and windowing sizes) for MLLM-based human activity recognition.}
  
  \vspace{-3pt}
  \label{fig:teaser}
}

\abstract{
Large Language Models (LLMs) have emerged as foundation models for IoT applications such as human activity recognition (HAR).
However, directly applying high-frequency and multi-dimensional sensor data, such as eye-tracking data, leads to information loss and high token costs.
To mitigate this, we investigate a visual prompting strategy that transforms sensor signals into data visualization images as an input to multimodal LLMs (MLLMs) using eye-tracking data.
We conducted a systematic evaluation of MLLM-based HAR across three public eye-tracking datasets using three visualization types of timeline, heatmap, and scanpath, under varying temporal window sizes.
Our findings suggest that visual prompting provides a token-efficient and scalable representation for eye-tracking data, highlighting its potential to enable MLLMs to effectively reason over high-frequency sensor signals in IoT contexts.
} 

\begin{document}
\maketitle

\section{Introduction}


In recent years, Large Language Models (LLMs) have been increasingly adopted as foundation models for IoT applications to understand human context and make autonomous decisions. Prior works have demonstrated grounding LLMs with diverse sensor data to support a wide range of IoT tasks, including human activity recognition (HAR)~\cite{xu2024penetrative, ji2024hargpt}, data sensemaking~\cite{li2025vital}, and health-related state inference from wearable sensing~\cite{liu2023large}. More recently, LLMs have been explored as reasoning agents for IoT scenarios~\cite{yang2024drhouse}. 
Collectively, these studies investigate and validate the feasibility of adopting LLMs as the foundation model for IoT ecosystems.
Unlike traditional machine learning approaches, integration of LLMs enables training-free IoT applications, bypassing the laborious effort for 
massive labeled data collection 
and provides generalization capabilities with the LLM's flexible world knowledge~\cite{wei2025one}.

However, significant challenges remain in utilizing LLMs in tasks with numeric, high-frequency time-series sensor data. 
First, LLM tokenizers are inherently designed for natural language and fail to capture patterns in continuous numerical
sequences, resulting in the loss of temporal correlations~\cite{spathis2024first}.
Second, high-frequency raw sensor data expands to an extremely large number of tokens, incurring infeasible computational cost and often prohibitive token costs. Importantly, the lengthy input triggers the ``lost in the middle'' problem~\cite{liu2024lost}, causing critical temporal patterns embedded in long sensor data to be overlooked.
To mitigate this, prior approaches have focused on aligning time-series modalities with LLMs, either by developing specialized tokenizers and encoders to handle numerical sequences efficiently~\cite{zhou2023one} 
or by training foundation models directly on large-scale time-series datasets such as inertial measurement unit (IMU) data~\cite{xu2021limu, zhao2025tartan}.
In particular, Yoon et al.~\cite{yoon2024my} utilized visual prompting\footnote{
In this paper, we use the term \textit{visual prompting} to denote the use of data visualization images as model inputs to guide a model with visual cues~\cite{yoon2024my}.
} to mitigate the problem, which transformed sensor data (e.g., accelerometer and physiological data) into visualizations such as line plots or spectrograms interpreted by multimodal LLMs (MLLMs). Their findings showed this approach outperformed text-based prompting in both HAR accuracy and token efficiency.

Meanwhile, eye-tracking serves as an aspiring sensing modality in intelligent systems, enabling diverse applications to infer user attention or cognition in pervasive environments such as smart homes and head-mounted displays~\cite{hansen2016wrist, barz2021automatic}.
Despite its utility, effectively grounding LLMs in eye-tracking data remains difficult due to its high-frequency, spatio-temporal nature, which requires combining 2D spatial coordinates with temporal dynamics.
At the same time, this characteristic allows diverse visualization techniques such as scanpaths~\cite{noton1971scanpaths} or heatmaps~\cite{mackworth1958eye}, which have been employed depending on specific analytical objectives.
In this work, we leverage these distinctive visual patterns inherent in eye-tracking visualizations to examine their effectiveness as visual prompts for multimodal LLMs. 
To validate this approach, we employ HAR, one of the key tasks in LLM-based IoT research~\cite{xu2024penetrative, ji2024hargpt}.
Through this experiment, we aim to 
demonstrate how different visual prompting strategies, such as visualization techniques and window sizes, can effectively represent the distinct behavioral patterns captured across multiple eye-tracking datasets.
Furthermore, we discuss the future development and broader applicability of this approach.


\section{Related Works}

\subsection{LLM-Based Human Activity Recognition}
\vspace{-1pt}
HAR has long been a central component in IoT and Cyber-Physical Systems (CPS), as it enables systems to interpret and respond to human behaviors from sensor data. Traditional approaches have focused on developing task-specific models using rule-based, machine learning, or neural networks trained on labeled datasets. More recently, a growing body of research has explored the use of LLMs as foundation models for HAR~\cite{ji2024hargpt, wei2025one, li2025sensorllm, hong2025llm4har,  xu2025exploring, yan2025large}, motivated by reduced training costs and potential generalizability~\cite{wei2025survey}.
One of the major examples is HARGPT~\cite{ji2024hargpt}, which demonstrated zero-shot HAR by feeding IMU data into the model, leveraging its internal world knowledge. 
A recent study also explored LLMs' potential in fine-grained hand gesture recognition, highlighting the importance of domain-specific adaptation and few-shot learning~\cite{xu2025exploring}. 

In LLM-based HAR tasks, sensor data are commonly provided directly within the in-context prompt without additional model training. Prompts often employ a specific expert persona (e.g., “You are an expert in human activity analysis”), provide data collection context such as device specifications and sampling rates, and include few-shot examples when necessary. The prompt also explicitly specifies the target task (e.g., identifying the activity performed based on the provided data).
When supplying time-series sensor data to LLMs via their context windows, the data are formatted in various ways, ranging from raw numerical strings~\cite{ji2024hargpt, hong2025llm4har}, statistical summaries~\cite{hong2025llm4har, yan2025large}, to high-level natural language descriptions of signal trends~\cite{ li2025sensorllm, yan2025large}.
Among these diverse input strategies, our work adopts the visual prompting approach of Yoon et al.~\cite{yoon2024my}, 
transforming sensor data into images to leverage the visual reasoning capabilities of MLLMs. Specifically, we apply this approach to eye-tracking data, which is two-dimensional and temporal data.

\vspace{-4pt}
\subsection{Eye-Tracking Visualization Techniques}
\vspace{-1pt}
Eye-tracking data can be visualized in different ways depending on the analytical purpose and the type of information to be revealed.
According to the taxonomy proposed by Blascheck et al.~\cite{blascheck2017visualization}, eye-tracking visualizations can be categorized into (i) Area-of-Interest (AOI)-based or (ii) point-based.
AOI-based visualizations focus on semantic information, analyzing the transition between predefined regions or objects~\cite{raiha2005static}. 
In contrast, point-based visualizations utilize the spatial and temporal information of recorded data points (e.g., $(x, y)$ coordinates) without requiring semantic annotations.
Due to the unique nature of eye-tracking data, many point-based visualizations rely on preprocessed eye-movement features, most notably fixations and saccades.
Fixations refer to relatively stable gaze periods lasting approximately 200--300\, ms, 
whereas saccades denote rapid eye movements between fixations~\cite{holmqvist2011eye}.
Feature extraction is commonly performed using velocity-based or dispersion-based methods, including I-VT, I-DT, and I-HMM~\cite{salvucci2000identifying}.





Within point-based visualizations, techniques can be differentiated based on their temporal and spatial emphasis.
\textbf{Timeline} plot represents temporal characteristics of gaze data by mapping gaze positions or derived features along a time axis. Typically, raw gaze coordinates are plotted as functions of time, enabling analysts to inspect temporal fluctuations in gaze behavior~\cite{goldberg2010visual}. More advanced variants additionally visualize fixations and saccades over time to explicitly encode attentional shifts~\cite{goldberg2010visual, grindinger2010group}.
\textbf{Heatmap} represents the spatial distribution of gaze~\cite{blascheck2017visualization} by overlaying gaze positions directly onto the stimulus to convey where viewers focused their attention~\cite{mackworth1958eye}. As reported by Bojko~\cite{bojko2009informative}, heatmaps can be further classified based on the level of aggregation, including representations of absolute raw gaze points and fixation-based heatmaps that encode fixation counts or durations.
\textbf{Scanpath} depicts gaze trajectories to represent the spatio-temporal sequence of visual exploration~\cite{noton1971scanpaths}. A common usage involves rendering fixations as nodes and saccades as connecting paths, where variables such as the circle radius are often mapped according to the fixation duration~\cite{scinto1986cognitive}. Additional visual encoding, such as gradient, can be employed to convey the temporal order of sequences~\cite{lankford2000gazetracker}.
\textbf{Space--time cube (STC)}
    extends the 2D spatial domain of a stimulus with a temporal dimension, resulting in a 3D representation of gaze~\cite{kurzhals2013space}. 

Building upon these established visualization techniques, we investigate their use as visual prompts, whether the techniques originally designed for human analysis can also support visual reasoning in MLLMs. 
Compared to our prior work~\cite{choi2025gaze2prompt}, we broaden the scope by considering a more diverse set of visualization techniques and varying temporal window sizes across multiple datasets.


\vspace{-4pt}
\section{Research Methodology}

\vspace{-1pt}

\subsection{Visualization Selection and Design}
\vspace{-1pt}
\begin{figure}
    \centering
    \includegraphics[width=\linewidth]{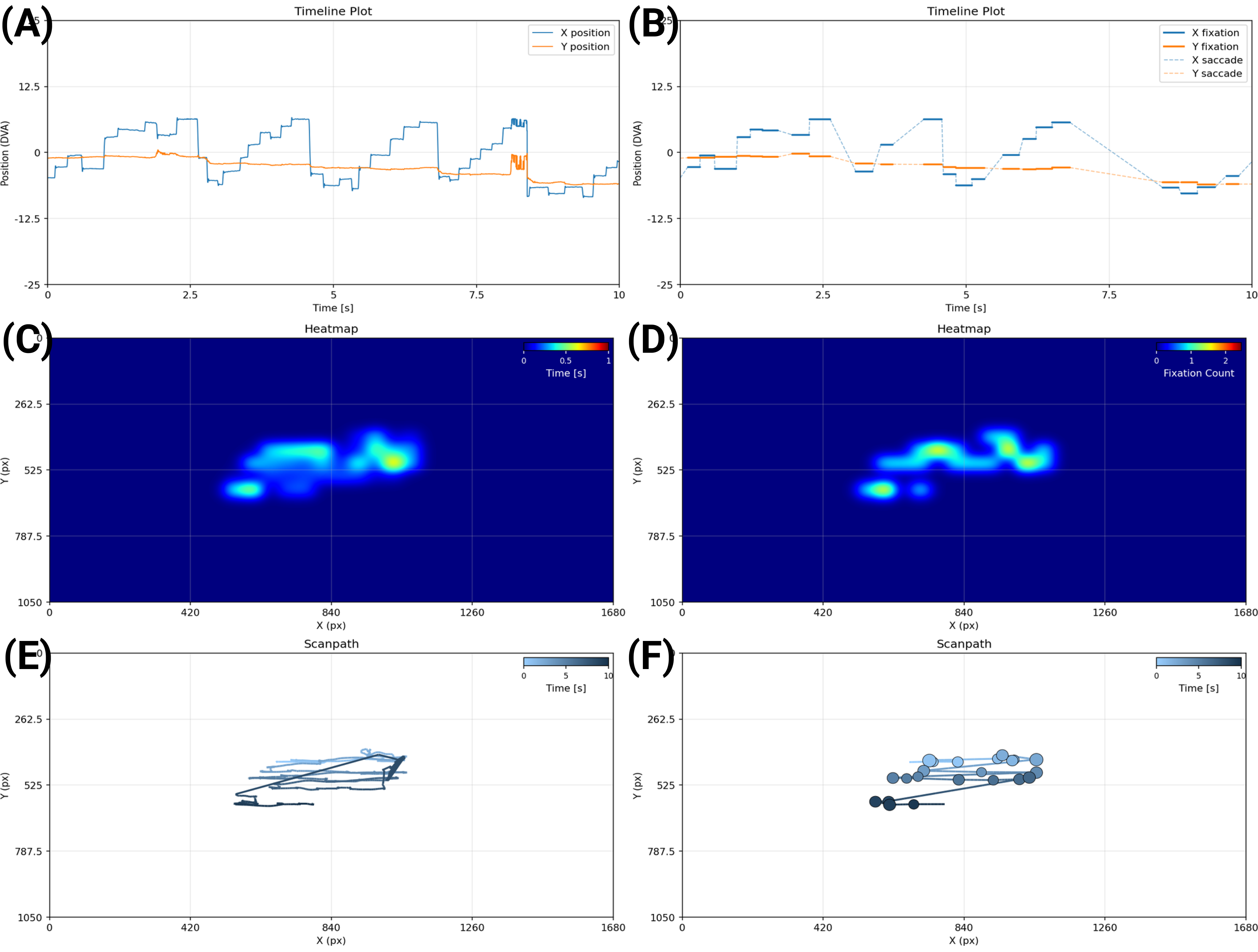}
    \vspace{-15pt}
    \caption{Visualizations used in this study. (A) Raw Timeline (B) Feature-based Timeline (C) Absolute Duration Heatmap (D) Fixation Count Heatmap (E) Raw Scanpath (F) Feature-based Scanpath.}
    
    \vspace{-18pt}
    \label{fig:visualization}
\end{figure}

In this study, we focus only on point-based eye-tracking visualization to accommodate general IoT scenarios where semantic AOIs are not readily defined. 
Specifically, we consider only 2D visualization techniques, excluding STCs.
To systematically evaluate the efficacy of visual prompting 
across different visualization techniques, we designed six distinct visualizations (Figure~\ref{fig:visualization}), categorized into three primary representations: (i) \textbf{Timeline}, (ii) \textbf{Heatmap}, and (iii) \textbf{Scanpath}. For each category, we compared raw data representations with feature-based representations derived from fixations and saccades extracted using the I-DT algorithm~\cite{salvucci2000identifying}.
While the dispersion thresholds were adaptively adjusted for each dataset, the minimum dwell time for fixation detection was fixed at 200\, ms. 

Regarding the visualization techniques, the Timeline plot using the \textit{raw data} (Figure~\ref{fig:visualization}A) represents $x$ and $y$ coordinates as distinct colored lines over time. For the \textit{feature-based} Timeline plot (Figure~\ref{fig:visualization}B), fixations are illustrated with thick solid lines, while saccades are depicted as dashed lines to distinguish gaze events. 
The Heatmaps (Figure~\ref{fig:visualization}C, D) were generated on a blue background by partitioning the screen into 50\, px grids and using matplotlib’s Gaussian interpolation for visual smoothing. 
Finally, for the Scanpaths (Figure~\ref{fig:visualization}E, F), a linear temporal gradient ranging from light blue to dark blue was applied, where the radius of each fixation point was mapped to its respective duration. All visualizations are formatted to $1024 \times 512$ pixels, which corresponds to two $512 \times 512$ unit tiles for the OpenAI api, resulting in a token count of 350 each when using GPT-5.1~\cite{openai_pricing}.
Also, we implemented two text-based prompting approaches for the baselines. \textit{Raw Text} represents the eye-tracking data as a sequence of $(x, y)$ coordinates formatted as comma-separated values. \textit{Feature Text} is a structured symbolic representation which encodes fixations as 
$F((x, y), t)$ and saccades as $S((x_{\text{start}}, y_{\text{start}}) \rightarrow (x_{\text{end}}, y_{\text{end}}), t)$, where $t$ denotes the duration.

\subsection{Datasets}

To assess the generalizability of this approach, we utilize three public eye-tracking datasets: \textit{GazeBase}~\cite{griffith2021gazebase}, \textit{SedentaryActivity}~\cite{srivastava2018combining}, and \textit{DesktopActivity}~\cite{lan2020gazegraph}. 
Each dataset comprises a distinct set of desktop-based activities, collected using either screen-based or egocentric wearable devices. 
A comprehensive summary of the dataset specifications is provided in Table~\ref{tab:datasets}.

\begin{table}[t]
\centering
\caption{Overview of the eye-tracking datasets used in this study.}
\vspace{-4pt}
\label{tab:datasets}
\setlength{\tabcolsep}{3pt}
\renewcommand{\arraystretch}{1.1}
\normalsize
\resizebox{\columnwidth}{!}{
\begin{tabular}{l|ccc}
 & \textit{GazeBase}&  \textit{SedentaryActivity}&  \textit{DesktopActivity}\\
\midrule
    \textbf{Sampling Rate} & 1000 Hz & 30 Hz & 30 Hz \\
    \textbf{Sensor Type} & Screen-based & Screen-based & Wearable \\
    \textbf{Device} & EyeLink 1000 & Tobii Pro X2-30 & Pupil Core \\
    \textbf{Activity Classes} & 6 & 8 & 6 \\
    \textbf{Participants} & 14 (Round 9) & 24 & 8 \\
    \textbf{Screen Spec} & 1680 $\times$ 1050 px & 24-inch monitor & 34-inch monitor \\
    \textbf{Coord. Unit} & dva (deg) & pixel & normalized (0-1) \\ 
\bottomrule
\end{tabular}%
}
\vspace{-16pt}
\end{table}

\textbf{\textit{GazeBase}}: 
This dataset provides longitudinal eye-tracking data collected at 1,000 Hz using EyeLink 1000~\cite{srresearch_eyelink1000}. 
It consists of six activities: \textit{Horizontal Saccade}, \textit{Random Saccade}, \textit{Fixation}, \textit{Reading}, \textit{Video Viewing}, and \textit{Gaze-driven Game}. The original dataset includes two video viewing activities; we utilized only one for distinct class separability. 
To ensure data stability, we used data from the final session (Round 9) collected from 14 participants.
For this dataset only, the \textit{Raw Text} representation was downsampled by 1/10 due to its high sampling rate, considering the model's context limit. 

\textbf{\textit{SedentaryActivity}}: 
Capturing naturalistic desktop behaviors, this dataset was recorded at 30 Hz using a Tobii Pro X2-30~\cite{tobii_x2_30} from 24 participants. 
It encompasses eight computing tasks: \textit{Read}, \textit{Watch}, \textit{Browse}, \textit{Search}, \textit{Play}, \textit{Interpret}, \textit{Debug}, and \textit{Write}. It contains coding-related tasks (\textit{Interpret}, \textit{Debug}, and \textit{Write}), and each activity is composed of three predefined sub-activities.
For this study, we utilized the raw pixel coordinates provided by the tracker.

\textbf{\textit{DesktopActivity}}: 
Unlike the previous datasets, this dataset was collected using a wearable egocentric device (Pupil Core~\cite{pupillabs_pupilcore}) at 30 Hz.
Data were gathered from eight participants performing six activities: \textit{Browse}, \textit{Play}, \textit{Read}, \textit{Search}, \textit{Watch}, and \textit{Write}.
Due to the absence of specific screen resolution metadata, we utilized the provided normalized coordinates ($0$ to $1$) for our analysis.

\subsection{Experimental Design}
We first investigated the impact of varying visual prompting strategies for HAR under both zero-shot and one-shot settings. 
For a one-shot setting, we constructed an example pool by randomly selecting 2, 3, and 1 participants for the datasets, respectively, in proportion to the number of participants in each dataset (14, 24, and 8). The remaining participants served as test data to prevent overlap between the example and test sets. In each experimental run, data from one randomly sampled participant within the example pool were used as the one-shot examples.
For evaluation, we generated 30 test cases per activity class.
Since the three datasets contain six, eight, and six activity classes, respectively, this resulted in 180, 240, and 180 test cases per dataset for each experimental condition.
We used the state-of-the-art MLLM \texttt{gpt-5.1-2025-11-13} via the OpenAI API with default inference settings. The \texttt{"detail": "high"} parameter was applied only to image inputs.

Furthermore, to evaluate the effect of window size, we conducted additional experiments on the \textit{SedentaryActivity} and \textit{DesktopActivity} datasets, varying the window size from 20 to 100 seconds in 10-second increments. The \textit{GazeBase} dataset was excluded from this analysis as its \textit{Fixations} activity segments are notably short, with an average duration of 14.7 seconds. In these experiments, we focused exclusively on one-shot settings, utilizing four feature-based representations: Timeline, Heatmap, Scanpath, and feature text as a baseline. The selection process for one-shot examples and test cases followed the same protocol as the primary experiment.

Our prompt design follows the structured framework recommended for IoT foundation models~\cite{wei2025survey}. 
In the system prompt, we defined the model’s role as a domain expert in eye-tracking activity recognition and instructed it to analyze either visual or numerical eye-tracking inputs. 
And the model was required to output a structured response containing the predicted activity label and a short explanation of the reason for its decision.
The user prompt is composed of five segments: \texttt{Instruction}, which reiterates the HAR task and expert role; \texttt{Activity Descriptions}, which provides informations of each activities and stimuli; \texttt{Context}, which includes sensor specifications such as sampling rate and window size; \texttt{Examples} (one-shot setting only), which present one representative visualization or text example per activity class; and \texttt{Question}, which contains the target instance to be classified. 
To mitigate ordering bias in language models~\cite{zheng2023judging}, we randomized the order of activity descriptions and example activities for every query. 

\section{Results}
\newcommand{\dataset}[1]{\makebox[1.75cm][c]{\textit{#1}}}
\newcolumntype{L}[1]{>{\raggedright\arraybackslash}p{#1}}
\newcommand{\sixpoint}{\fontsize{7pt}{6pt}\selectfont}
\newcommand{\overlapnote}[1]{\makebox[0pt][c]{\sixpoint #1}}
\begin{table}[t]
\caption{HAR accuracy and token consumption across experimental conditions (10s window). \textbf{Bold} indicates best performance per dataset, and \textcolor{red}{red} denotes performance below the textual baseline. Multipliers ($\times\uparrow$) show the relative increase in token usage of textual prompts compared to visual prompts.}

\vspace{-4pt}
\centering
\small
\begin{tabular}{L{0.75cm} L{0.35cm} c c c c c c}
\toprule
\noalign{\vskip -0.3ex}
 &  & \multicolumn{2}{c}{\dataset{GazeBase}}
   & \multicolumn{2}{c}{\dataset{SedentaryActivity}}
   & \multicolumn{2}{c}{\dataset{DesktopActivity}} \\
\noalign{\vskip -0.3ex}
\cmidrule(lr){3-4}
\cmidrule(lr){5-6}
\cmidrule(lr){7-8}
\noalign{\vskip -0.3ex}
 &  & 0-shot & 1-shot & 0-shot & 1-shot & 0-shot & 1-shot
 \\
\noalign{\vskip -0.3ex}
\midrule
\noalign{\kern-\aboverulesep}
\rowcolor{gray!12}
\textit{Accuracy} & & & & & & & \vspace{3pt}\\
\multirow{2}{*}{\textbf{Timeline}}
 & Raw   & 0.500  &\textbf{0.811}  & \textcolor{red}{0.096}  &0.283  & 0.200  &0.257  \\
 & Feat  & \textcolor{red}{0.450}  &\textcolor{red}{0.772}  & \textcolor{red}{0.067}  &\textcolor{red}{0.246}  & 0.200  &0.261  \\
\midrule
\multirow{2}{*}{\textbf{Heatmap}}
 & Raw   & 0.578  &\textbf{0.811}  & \textcolor{red}{0.138}  &\textbf{0.317}  & \textcolor{red}{0.183}  &\textbf{0.311}  \\
 & Feat  & 0.644  &\textcolor{red}{0.739}  & \textcolor{red}{0.129}  &0.304  & 0.217  &0.278  \\
\midrule
\multirow{2}{*}{\textbf{Scanpath}}
 & Raw   & 0.506  &0.694  & 0.183  &0.267  & 0.200  &0.300  \\
 & Feat  & 0.544  &\textbf{0.811}  & \textcolor{red}{0.138}  &\textcolor{red}{0.271}  & \textcolor{red}{0.194}  &0.272  \\
\midrule
\multirow{2}{*}{Text}
 & Raw   & 0.378  &0.539  & 0.154  &0.242  & 0.189  &0.211  \\
 & Feat  & 0.533  &0.794  & 0.167  &0.300  & 0.217  &0.244  \\
\midrule
\noalign{\kern-\aboverulesep}
\rowcolor{gray!12}
\multicolumn{3}{l}{\textit{Number of input tokens}}& & & & & \vspace{3pt}\\
\multicolumn{2}{l}{\textbf{Visual prompt}} & 1086 & 3247  & 1192 & 4061  & 1106 & 3259\\
\midrule
\multicolumn{2}{l}{Raw text}      & 10748 & 70818  & 2699 & 17678  & 3755 & 21814 \\
\multicolumn{2}{l}{ }      
& \overlapnote{(9.9$\times\uparrow$)}& \overlapnote{(21.8$\times\uparrow$)}  & \overlapnote{(2.3$\times\uparrow$)} & \overlapnote{(4.4$\times\uparrow$)}  & \overlapnote{(3.4$\times\uparrow$)} & \overlapnote{(6.7$\times\uparrow$)}  \vspace{1pt}\\
\multicolumn{2}{l}{Feature text}  & 1616 & 6828  & 1744 & 9131  & 1617 & 6412  \\
\multicolumn{2}{l}{ }      
& \overlapnote{(1.5$\times\uparrow$)}& \overlapnote{(2.1$\times\uparrow$)}  & \overlapnote{(1.46$\times\uparrow$)} & \overlapnote{(2.2$\times\uparrow$)}  & \overlapnote{(1.5$\times\uparrow$)} & \overlapnote{(2.0$\times\uparrow$)}  \\
\bottomrule
\end{tabular}
\label{tab:results}
\vspace{-18pt}
\end{table}

\subsection{Experiment 1: Impact of Visualization Techniques}
Table~\ref{tab:results} summarizes the HAR accuracy and input token consumption for the various prompting strategies across three datasets. 
On the \textit{GazeBase} dataset (six classes), the model achieved its highest accuracy of 0.811 under several visual prompting conditions.
In contrast, performance on the \textit{SedentaryActivity} (eight classes) and the \textit{DesktopActivity} (six classes) datasets was substantially lower, with the highest accuracies of only 0.317 and 0.311, making it difficult to identify clear performance differences across conditions.
These results suggest that a fixed 10-second window is insufficient to capture the activity patterns in these datasets, motivating our subsequent analysis with varying window sizes in Experiment 2.

\begin{figure}
    \centering
    \includegraphics[width=\linewidth]{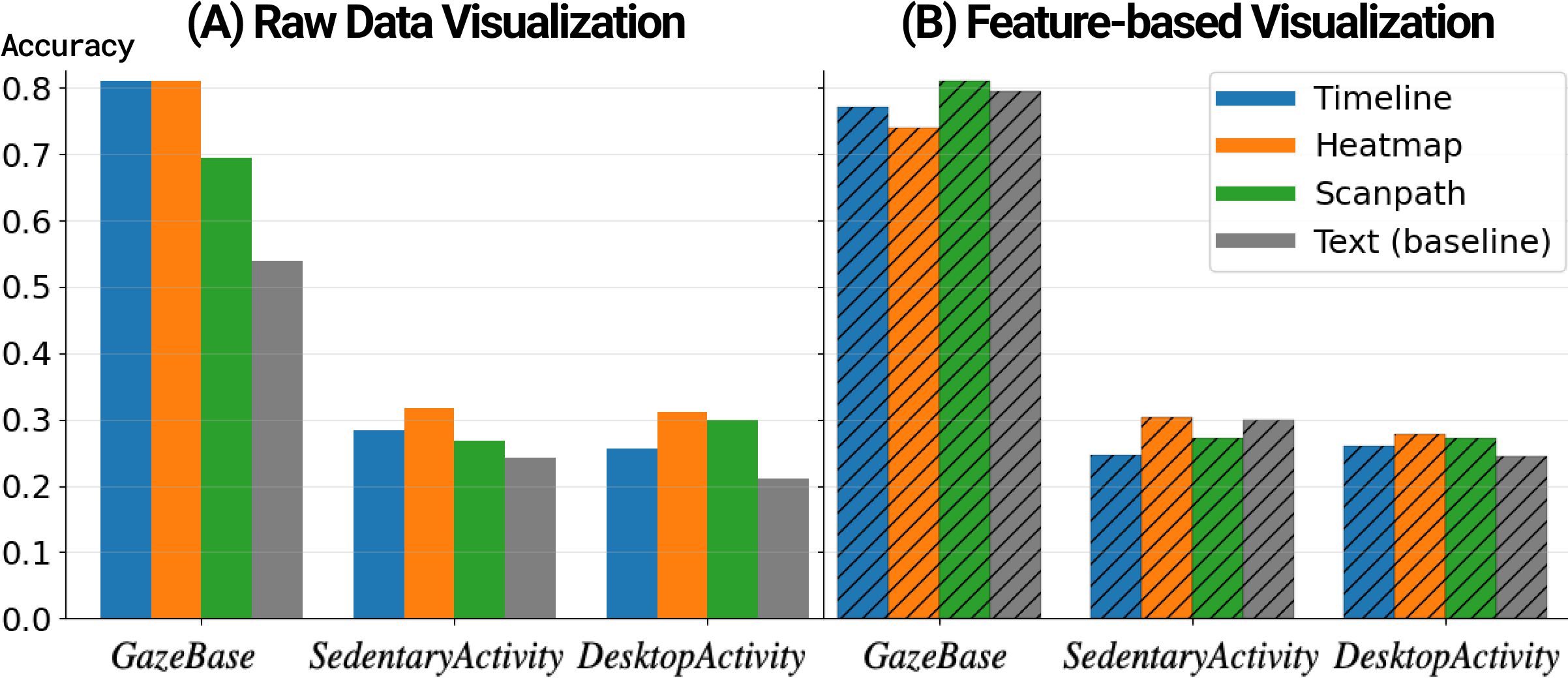}
    \vspace{-12pt}
    \caption{One-shot setting accuracy comparison between visualization techniques across three datasets.}
    \label{fig:oneshot}
    \vspace{-12pt}
\end{figure}

Compared to textual prompt baselines, visual prompts generally achieved higher performance with raw data.
In the one-shot setting, visual prompts using raw data achieved higher accuracy than textual prompts across all datasets and visualization types (Figure~\ref{fig:oneshot}(A)), and this performance gap is even more apparent for the \textit{GazeBase} dataset, where the Timeline and Heatmap (0.811) outperformed raw text (0.539). In the zero-shot setting, visual prompts with raw data continued to outperform textual prompts, except in a few cases, such as the Timeline and Heatmap on \textit{SedentaryActivity} and the Heatmap on \textit{DesktopActivity}. 
Notably, across all three datasets in the one-shot setting, the Heatmap with raw data consistently achieved the highest performance (Figure~\ref{fig:oneshot}(A)).

However, the advantage of visual prompting does not consistently hold when compared with feature-based textual representations. As shown in Figure~\ref{fig:oneshot}(B), the feature text prompt exhibited performance comparable to visual prompts across three datasets.
In \textit{SedentaryActivity}, for instance, the feature text prompt outperformed most visual prompting conditions both in zero-shot and one-shot settings. Similarly, in \textit{GazeBase} one-shot setting, only the Scanpath surpassed the feature text accuracy.
These results indicate that explicit gaze features are effective even in text form, compared to raw textual sequences.
From another perspective, results on \textit{GazeBase} show that raw Timeline and Heatmap visualizations achieve accuracy comparable to the feature text prompt. This suggests that in some cases, appropriate visual abstractions can bypass the need for explicit feature extraction.

When comparing raw data and feature-based representations within the same visualization type, their accuracies exhibited distinct differences.
For the Timeline, using raw data generally yielded higher accuracy than feature-based visualizations, except for \textit{DesktopActivity} one-shot setting. 
Similarly, for Heatmap, raw data likewise outperformed feature-based representations under the one-shot setting.
In contrast, the Scanpath results varied across datasets. For \textit{GazeBase}, the feature-based representation outperformed raw data in both the zero-shot and one-shot settings, whereas the opposite trend was observed for \textit{SedentaryActivity} and \textit{DesktopActivity}. We attribute this pattern to differences in activity compositions across datasets, suggesting that certain activities may benefit more from the explicit encoding of fixations and saccades.
In terms of token consumption, visual prompting offers a substantial advantage. As detailed in Table~\ref{tab:results}, visual prompts consumed fewer tokens than textual prompts. The raw text baseline required between 2.3$\times$ and 21.8$\times$ more tokens than visual prompts to represent the same 10-second window.
Even the more condensed feature text consumed 1.46$\times$ to 2.2$\times$ more tokens.
These findings highlight visual prompting as a cost-effective approach for processing high-frequency eye-tracking data with MLLMs.

\subsection{Experiment 2: Impact of Window Size}
\label{exp2}
\begin{figure}
    \centering
    \includegraphics[width=\linewidth]{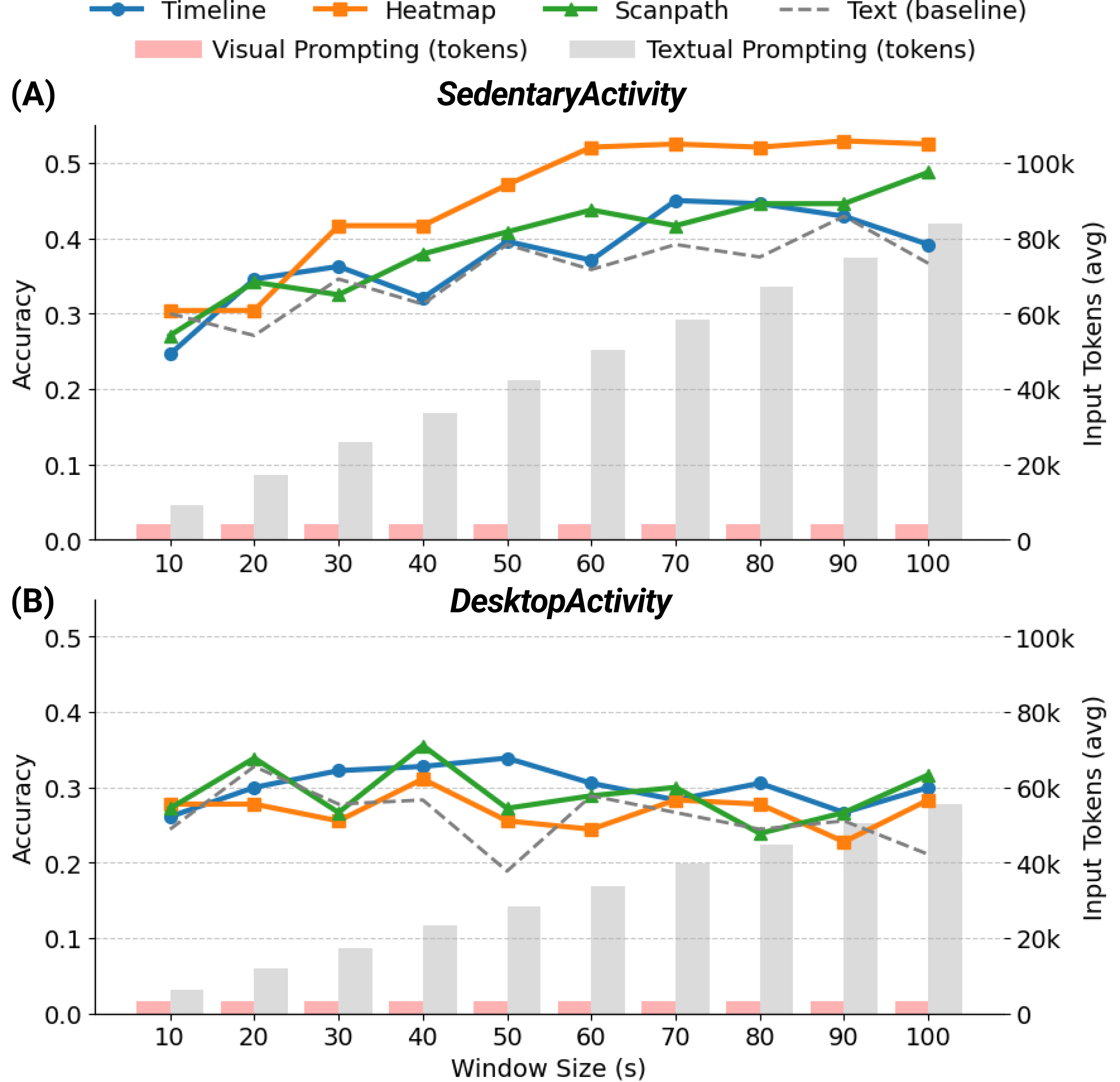}
    \vspace{-12pt}
    \caption{Comparison of HAR accuracy and token consumption across varying temporal window sizes (10-second to 100-second) for the (A) \textit{SedentaryActivity} and (B) \textit{DesktopActivity} datasets. The line plots illustrate the accuracy trends for each visual prompting strategy (Timeline, Heatmap, Scanpath, and the textual baseline). The grouped bar charts represent the corresponding input tokens for each condition.}
    \label{fig:window}
    \vspace{-12pt}
\end{figure}

Figure~\ref{fig:window} presents the classification accuracy and token consumption results for the \textit{SedentaryActivity} and \textit{DesktopActivity} datasets across temporal window sizes ranging from 10 to 100 seconds. 
In \textit{SedentaryActivity}, accuracy generally improved as the window size increased. The Heatmap achieved the highest accuracy across nearly all window sizes except for the 20-second condition. Performance for the Heatmap showed signs of saturation beyond 60 seconds, maintaining a stable range between 0.521 and 0.529. 
The performance ranking remained stable (Heatmap, Scanpath, Timeline, and the textual baseline) for window sizes of 40 seconds or longer, except at 70 seconds, where Timeline and Scanpath switched orders. Unlike the Heatmap, the Scanpath visualization exhibited a steady upward trend without reaching a clear saturation point, peaking at 0.488 at 100-second. Only the Timeline exhibited a peak in performance and reached an accuracy of 0.450 at 70 seconds before declining to 0.392 at 100 seconds. Although the textual baseline also showed improvement with larger windows, its maximum accuracy (0.429 at 90 seconds) remained lower than the visual prompts.
Given the eight activity classes in this dataset, the highest accuracy of 0.529 indicates that the model can capture relevant behavioral patterns when provided with sufficient temporal context.
Taken together, the results suggest that extended temporal windows combined with appropriate visual representations facilitate more effective utilization of gaze dynamics in \textit{SedentaryActivity}.

In contrast, \textit{DesktopActivity} did not exhibit a clear accuracy pattern related to window size. The accuracy remained around 0.3 under all conditions. The highest performances were recorded at different intervals: 0.339 at a 50-second window for the Timeline, 0.356 at a 40-second window for the Scanpath, and 0.311 at a 40-second window for the Heatmap. In particular, the Heatmap showed relatively poor performance compared to other visual prompts in this dataset, which contradicts the results seen in \textit{SedentaryActivity}. Furthermore, no distinct performance gap was observed between visual and textual prompting strategies.
These results suggest that neither window size nor prompt modality strongly influences HAR accuracy in this dataset.

Regarding token usage, the visual prompts offer a clear advantage over the textual prompts. 
Token consumption for textual prompts scales linearly with temporal duration, whereas visual prompts maintain a constant token count regardless of window size. 
In the \textit{SedentaryActivity} dataset, visual prompts consistently utilized 4,061 tokens. 
At the 100-second window, the textual prompt required 83,901 tokens, corresponding to a $20.7\times$ increase over the visual prompt. 
Similarly, for \textit{DesktopActivity}, visual prompts consumed a fixed 3,259 tokens, whereas the textual baseline reached 55,481 tokens at 100 seconds, representing a $17.0\times$ increase. 
These results demonstrate that visual prompting provides a scalable approach for analyzing long-duration sensor data without the proportional increase in computational cost associated with textual representations.

\section{Discussions}



\begin{figure}
    \centering
    \includegraphics[width=\linewidth]{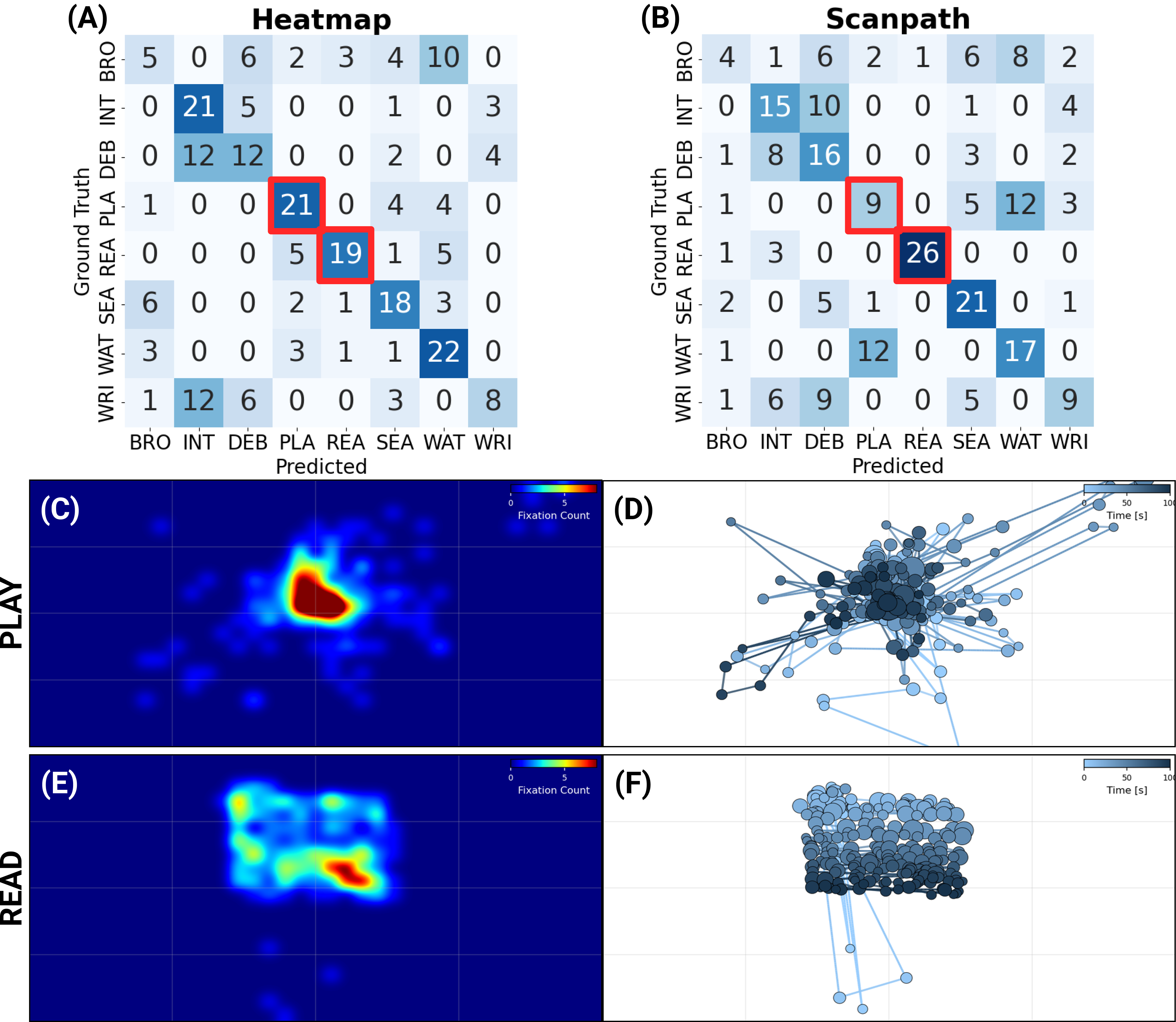}
    \vspace{-12pt}
    \caption{Confusion matrices for (A) Heatmap and (B) Scanpath visual prompting, alongside representative examples for PLAY (C, D) and READ (E, F) from the \textit{SedentaryActivity} 100-second window experiment. Activity names in the matrices are abbreviated to the first three characters. Titles, axis titles and ticks for the examples (C--F) are omitted to enhance visual clarity.}
    \label{fig:placeholder}
    \vspace{-12pt}
\end{figure}

A primary advantage of visual prompting lies in its scalability and token efficiency.
In the \textit{GazeBase} dataset, which consists of experimental tasks such as \textit{Horizontal Saccade} and \textit{Fixation}, a 10-second window was sufficient for the model to distinguish between classes. 
In contrast, the \textit{SedentaryActivity} dataset comprises more complex and naturalistic activities (e.g., \textit{Search}, \textit{Debug}), where applying larger windows to include more behavioral information proved more effective.
This observation aligns with the findings of Lan et al.~\cite{lan2020gazegraph}, who reported that recognition performance increased as the sensing window size grew from 10 to 30 seconds in their graph-based gaze representation learning framework.
While providing extended temporal context is beneficial for these complex activities, textual prompting suffers from the “lost in the middle” problem~\cite{liu2024lost} and reduced computational efficiency. 
Visual prompting can overcome these limitations by maintaining a constant token cost within a fixed canvas, regardless of the data length. 
In particular, the Timeline representations can offer broader extensibility, as their temporal structure can be readily adapted to other time-series sensor modalities (e.g., IMU or physiological signals).

However, this efficiency involves a trade-off with information density. As observed in the Timeline plots for the \textit{SedentaryActivity} dataset, accuracy may decline when the temporal window exceeds a certain threshold due to visual overcrowding and subsequent information loss.
This suggests that maintaining an optimal information density is a critical factor in designing effective visual prompts for high-frequency sensor data.
Finally, the visual prompting approach did not yield significant performance gains in the \textit{DesktopActivity} dataset, which may be caused by its unique characteristics such as the use of a egocentric sensor or potential overfitting to the provided examples, which necessitates further investigation.

Furthermore, there was no one-size-fits-all visualization technique.
Although we observed the strengths of Heatmap in both Experiment 1 and 2, the effectiveness of a specific visualization appears to be highly dependent on the behavioral characteristics of the target activity.
In the 100-second window experiment for the \textit{SedentaryActivity} dataset (Figure~\ref{fig:placeholder}), the \textit{Play} activity was more effectively classified using the Heatmap (Figure~\ref{fig:placeholder}(C)) than the Scanpath (Figure~\ref{fig:placeholder}(D)) (21 vs. 9). In contrast, the \textit{Read} activity showed higher accuracy with the Scanpath (Figure~\ref{fig:placeholder}(F)) compared to the Heatmap (Figure~\ref{fig:placeholder}(E)) (26 vs. 19). 
This contrast indicates that different activities emphasize different aspects of gaze behavior. For \textit{Play}, the spatial distribution and density of attention are more informative, while for \textit{Read}, the temporal ordering of gaze movements plays a more central role. These results highlight the importance of choosing visualizations for MLLMs that align with the task-specific behavioral structure.
Our findings can contribute to the design of an autonomous agent capable of selecting the most representative eye-tracking visualization for a given context.
Additionally, combining multiple visualization techniques to integrate their strengths may offer improvements and warrants further investigation.

While the HAR accuracy of visual prompting currently remains lower than that of conventional approaches~\cite{lan2020gazegraph}, 
it offers a training-free alternative that leverages the pre-existing world knowledge and visual reasoning capabilities of MLLMs~\cite{wei2025survey}.
By utilizing visual encoders to interpret gaze patterns, this method provides a novel framework for utilizing data visualization as a bridge between human context and foundation models in the IoT ecosystem.
Beyond HAR, this framework may support higher-level inference about user attention or task engagement, which are central to pervasive systems in ubiquitous computing scenarios. Rather than limiting the model to predicting predefined activity labels, visual prompts could support more flexible reasoning, such as inferring whether a user is focused, distracted, or searching for a component. In our study, we evaluated visual prompting on three datasets covering a diverse range of activities; however, future investigation is needed to assess its applicability in more open-ended real-world IoT scenarios. Taken together, our findings suggest that visual prompting may serve as a flexible intermediate representation for context-aware reasoning without task-specific retraining, while its generalizability requires further empirical validation.
\section{Conclusion}

In this study, we investigated the efficacy of visual prompting for eye-tracking-based HAR using MLLMs.
We designed and evaluated three visualizations across three public datasets and multiple temporal window sizes.
Our findings provide a valuable insight that visual prompting serves as a token-efficient and scalable alternative to textual prompting for MLLMs, especially for long duration eye-tracking data.
We discussed the importance of task-specific visualization selection and the applicability of sensor data visualization in foundation model contexts. 
This study has several limitations, including reliance on closed-source models, the lack of direct comparisons with conventional HAR methods, and its focus being limited to HAR evaluation within IoT tasks.
Finally, future work will explore real-world application scenarios that leverage visual prompting as a interface for MLLM-based IoT systems.

\section*{Supplemental Materials}
\label{sec:supplemental_materials}
Supplemental materials\footnote{Available at \href{https://eyetrackingvisualprompts.github.io}{https://eyetrackingvisualprompts.github.io}.} provide all visual prompts used in the experiments to support further analysis and reproducibility.

 \bibliographystyle{ieeetr}
 \bibliography{template}

\end{document}